\documentclass[prl,10pt,letterpaper,twocolumn,showpacs]{revtex4}

\usepackage[dvipdfm]{graphicx}
\usepackage{times}
\usepackage{amsmath}
\usepackage{amssymb}
\usepackage[dvipdfm]{hyperref}

\begin{document}

\title{Splitting of quantum information in traveling wave fields\\
using only linear optical elements}
\author{W. B. Cardoso}
\affiliation{Instituto de F\'{\i}sica, Universidade Federal de Goi\'{a}s, 74.001-970, Goi%
\^{a}nia - GO, Brazil}
\author{N. G. de Almeida}
\affiliation{Instituto de F\'{\i}sica, Universidade Federal de Goi\'{a}s, 74.001-970, Goi%
\^{a}nia - GO, Brazil}
\author{A. T. Avelar}
\affiliation{Instituto de F\'{\i}sica, Universidade Federal de Goi\'{a}s, 74.001-970, Goi%
\^{a}nia - GO, Brazil}
\author{B. Baseia}
\affiliation{Instituto de F\'{\i}sica, Universidade Federal de Goi\'{a}s, 74.001-970, Goi%
\^{a}nia - GO, Brazil}
\pacs{03.67.Mn}

\begin{abstract}
In this brief report we present a feasible scheme to split quantum
information in the realm of traveling waves. An oversimplified scheme is
also proposed for the generation of a class of W states useful for perfect
teleportation and superdense coding. The scheme employs only linear optical
elements as beam splitters and phase shifters, in addition to photon
counters and one-photon sources. It is shown that splitting of quantum
information with high fidelity is possible even including inefficiency of
the detectors and photoabsorption of the beam splitters.
\end{abstract}

\maketitle


Entanglement is a commonplace of remarkable applications of Quantum
Mechanics, such as quantum computation \cite{Chuang}, superdense code \cite%
{Bennett92}, quantum teleportation \cite{Bennett93}, quantum communication
via teleportation \cite{Cirac95,Brassard98}, one-way quantum computation 
\cite{Raussendorf01_2}, quantum metrology \cite{Milburn2} and so on. A state
describing $N$ subsystems is entangled when it cannot be factorized into a
product of $N$ states, each one concerning with a subsystem. In this
respect, the $N$ subsystems are no longer independent, in spite of being
spatially separated. As a consequence, a measurement upon one of them not
only gives information about the other, but also provides possibilities of
manipulating it \cite{Horodecki09}.

There are various types of entangled states and classifying them is an
arduous task, specially for multipartite systems \cite{Leuchs}. However,
Bennett \textit{et al.} \cite{Bennett00} shed light in this question through
the use of the local operations and classical communication (LOCC) to define
classes of equivalence in the set of entangled states, i.e, two entangled
states belong to the same class of equivalence if one of them can be
obtained from the other with certainty by means of LOCC. According to this
criterion all bipartite pure-state entanglements are equivalent to that of
the EPR type $(|00\rangle +|11\rangle )/\sqrt{2}$ \cite{Einstein35}.
Concerning with tripartite states, in Ref. \cite{DurPRA00} it was shown, via
stochastic LOCC, that there are two genuine entanglement of tripartite
systems: GHZ ($(|000\rangle \pm |111\rangle )/\sqrt{2}$) and W ($%
(|100\rangle +|010\rangle +|001\rangle )/\sqrt{3}$) states, i.e., W (GHZ)
states cannot be converted into GHZ (W) states under stochastic LOCC.
Although the canonical W states cannot be used for perfect teleportation and
superdense coding, Ref. \cite{AgrawalPRA06} introduced the following class
of W states, suitable for these tasks, 
\begin{align}
|W_{\zeta }\rangle _{123}& =\frac{1}{\sqrt{2+2\zeta }}\left( |001\rangle
_{123}+\sqrt{\zeta }e^{i\gamma }|010\rangle _{123}\right.  \notag \\
& +\left. \sqrt{\zeta +1}e^{i\delta }|100\rangle _{123}\right) ,  \label{Ws}
\end{align}%
where $\zeta $ is a real number and $\gamma $ and $\delta $ are phases.
Latter on, in Ref. \cite{LiuJPA07} the W states were generalized to
multi-qubit and multi-particle systems with higher dimension. Recently, this
class of W states has called attention of researchers, as exemplified in
Ref. \cite{Becerra-CastroJPB08}, in the context of QED-cavity. A procedure
to split quantum information via W states was presented in Ref. \cite%
{ZhengPRA06}. This scheme relies on the following steps: i) the use of a
three qubit W state previously prepared and shared by Alice, Bob, and
Charlie; ii) a fourth qubit prepared in an unknown state (whose information
we want to split) in Alice's ownership; iii) a Bell-state measurement done
by Alice and informing her result to Bob and Charlie; iv) the agreement
between Bob and Charlie to send their particles, one to the other. After
these steps, the splitting quantum state shared by Bob and Charlie can be
reconstructed after an appropriate rotation.

In this brief report we show how to split a quantum state in the
domain of running wave fields. Our scheme is experimentally feasible: it
makes use of only linear optical elements, as beam splitters and phase
shifters, plus photodetectors and one-photon sources. A very attractive
scheme from the experimental point of view for generating the class of W
states given by Eq.(\ref{Ws}) is also presented. A W state of this class,
required to split the quantum information, composes the nonlocal channel
shared by Alice, Bob, and Charlie. Generation of W states in scenarios
different from traveling waves has been proposed in Ref. \cite{generation}.
The unknown quantum state used in our protocol to split the quantum
information is
\begin{equation}
\left\vert \psi \right\rangle _{a}=C_{0}\left\vert 0\right\rangle
_{a}+C_{1}\left\vert 1\right\rangle _{a}\text{ },  \label{IN}
\end{equation}%
where $C_{0}$ and $C_{1}$ are coefficients that obey $\left\vert
C_{0}\right\vert ^{2}+\left\vert C_{1}\right\vert ^{2}=1$. In Fig. 1 we show
the experimental setup corresponding to our proposal.

\begin{figure}[t]
\centering\includegraphics[width=6cm]{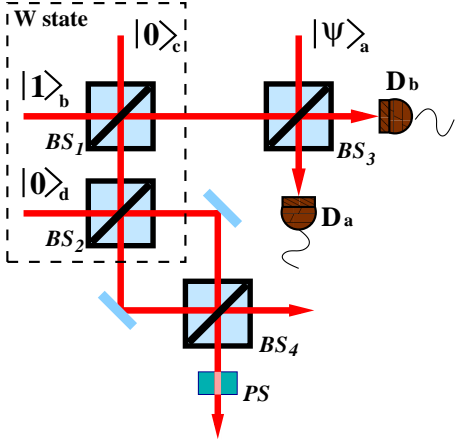}
\caption{(Color online) Scheme of the experimental setup required for splitting the quantum
information. The region in the dashed line consists in the W states
preparation. The $BS_{i}$, with $i=1,2,3,4$, are 50/50 beam splitters, PS is
the phase shift of $\protect\pi /2$, and $D_{a}$ ($D_{b}$) is the
photo-detector of the mode a (b).}
\end{figure}

\emph{W state generation} - To engineer the desired W state we employ two $%
50/50$ beam splitters, as shown in the dashed region of Fig. 1. The initial
state is given by $\left\vert \psi \right\rangle _{bcd}=\left\vert
1\right\rangle _{b}\left\vert 0\right\rangle _{c}\left\vert 0\right\rangle
_{d}$. After the interaction between modes $b-c$ in the $BS_{1}$ and modes $%
c-d$ in the $BS_{2}$, the state of the three qubits is%
\begin{equation}
\left\vert \psi \prime \right\rangle _{bcd}=\frac{1}{\sqrt{2}}\left\vert
1\right\rangle _{b}\left\vert 0\right\rangle _{c}\left\vert 0\right\rangle
_{d}+\frac{i}{2}\left\vert 0\right\rangle _{b}\left\vert 1\right\rangle
_{c}\left\vert 0\right\rangle _{d}-\frac{1}{2}\left\vert 0\right\rangle
_{b}\left\vert 0\right\rangle _{c}\left\vert 1\right\rangle _{d}.  \label{W}
\end{equation}%
Note that the W state given by Eq.(\ref{W}) is already of that class
which leads to perfect teleportation and superdense coding
with $\zeta =1$, $\gamma =3\pi /2$, and $\delta =\pi /2$ \cite{AgrawalPRA06}.

\emph{Quantum information splitting}\textit{\ }- To start our protocol for
splitting the information, the state given by Eq.(\ref{IN}) is sent to Alice,%
\textbf{\ }who shares with Bob and Charlie an entangled state of the W type
given by Eq.(\ref{W}). The state of the whole system reads, 
\begin{align}
\left\vert \phi \right\rangle _{abcd}& =\frac{1}{2}\left( \sqrt{2}%
C_{0}\left\vert 0,1,0,0\right\rangle _{abcd}+iC_{0}\left\vert
0,0,1,0\right\rangle _{abcd}\right.  \notag \\
& -C_{0}\left\vert 0,0,0,1\right\rangle _{abcd}+\sqrt{2}C_{1}\left\vert
1,1,0,0\right\rangle _{abcd}  \notag \\
& +\left. iC_{1}\left\vert 1,0,1,0\right\rangle _{abcd}-C_{1}\left\vert
1,0,0,1\right\rangle _{abcd}\right) .
\end{align}%
Alice now accomplishes a joint measurement on her qubits. The corresponding
Bell-states are 
\begin{eqnarray}
|\Psi ^{(\pm )}\rangle _{ab} &=&\frac{1}{\sqrt{2}}\left( \left\vert
0,1\right\rangle _{ab}\pm i\left\vert 1,0\right\rangle _{ab}\right) , \\
|\Phi ^{(\pm )}\rangle _{ab} &=&\frac{1}{\sqrt{2}}\left( \left\vert
1,1\right\rangle _{ab}\pm i\left\vert 0,0\right\rangle _{ab}\right) .
\end{eqnarray}%
After the Alice's measurement, the state of the particles $c$ and $d$ \ on
Charlie and Bob hands, respectively, collapses onto one of the entangled
states appearing below (up to normalization) 
\begin{eqnarray}
&&\left. |\Psi ^{\left( +\right) }\rangle _{ab}\left[ \sqrt{2}%
C_{0}\left\vert 0,0\right\rangle _{cd}+C_{1}\left( \left\vert
1,0\right\rangle _{cd}+i\left\vert 0,1\right\rangle _{cd}\right) \right]
\right.  \notag \\
&+&|\Psi ^{\left( -\right) }\rangle _{ab}\left[ \sqrt{2}C_{0}\left\vert
0,0\right\rangle _{cd}-C_{1}\left( \left\vert 1,0\right\rangle
_{cd}+i\left\vert 0,1\right\rangle _{cd}\right) \right]  \notag \\
&+&|\Phi ^{\left( +\right) }\rangle _{ab}\left[ C_{0}\left( \left\vert
1,0\right\rangle _{cd}+i\left\vert 0,1\right\rangle _{cd}\right) +\sqrt{2}%
C_{1}\left\vert 0,0\right\rangle _{cd}\right]  \notag \\
&+&\left. |\Phi ^{\left( -\right) }\rangle _{ab}\left[ C_{0}\left(
\left\vert 1,0\right\rangle _{cd}+i\left\vert 0,1\right\rangle _{cd}\right) -%
\sqrt{2}C_{1}\left\vert 0,0\right\rangle _{cd}\right] \right. .
\end{eqnarray}

\begin{figure}[t]
\includegraphics[width=7cm]{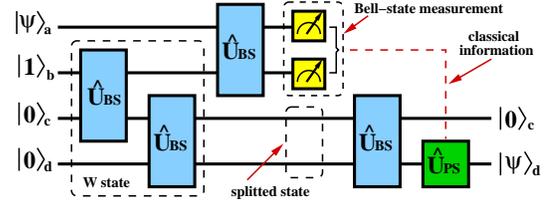}
\caption{(Color online) Schematic diagram of the quantum circuit. The $\widehat{U}_{BS}$
are the beam-splitter operator and $\widehat{U}_{PS}$ is the phase-shift
operator.}
\label{circuit}
\end{figure}

Note that the two components are not symmetric: while one of them
is a Bell state, the other one is a product state. The
reconstruction of the state can be done provided that Bob and Charlie
collaborate with each other. The $BS_{4}$, shown in Fig. 1, is used to
decouple the states corresponding to modes $c-d$, as shown in the following.
Fig. \ref{circuit} shows the schematic circuit to generate the W state and
to split quantum information.

\emph{Bell-state measurement} - The Bell-state measurement carried out by
Alice occurs in the $BS_{3}$ and in the photodetectors on modes $a$ and $b$.
After this joint measurement the Bell-state evolves to%
\begin{eqnarray}
|\Psi ^{(\pm )}\rangle _{a,b} &\rightarrow &\left\{ 
\begin{array}{c}
|1,0\rangle _{a,b}~~~~if~~~(+) \\ 
|0,1\rangle _{a,b}~~~~if~~~(-)%
\end{array}%
\right. ,  \label{bell1} \\
|\Phi ^{(\pm )}\rangle _{a,b} &\rightarrow &\left\{ |0,0\rangle
_{a,b},~~|2,0\rangle _{a,b},~~or~~|0,2\rangle _{a,b}\right. .  \label{bell2}
\end{eqnarray}%
Therefore, we can discern $|\Psi ^{(+)}\rangle _{a,b}$ from $|\Psi
^{(-)}\rangle _{a,b}$\ by detecting either $|1,0\rangle _{a,b}$ or $%
|0,1\rangle _{a,b}$ on detectors $D_{a}$ and $D_{b}$, as shown in Fig. 1.
Next, assuming for a moment that the detection corresponds to the states $%
|\Psi ^{(\pm )}\rangle _{a,b}$, the state of the particles $c$ and $d$ can
be written as%
\begin{equation}
N(\sqrt{2}C_{0}\left\vert 0,0\right\rangle _{cd}\pm C_{1}\left( \left\vert
1,0\right\rangle _{cd}+i\left\vert 0,1\right\rangle _{cd}\right) ),
\end{equation}%
where $N$ stands for normalization. Now, after the interaction on $BS_{4}$
the states corresponding to modes $c$ and $d$ decouple in the form 
\begin{equation}
|0\rangle _{c}(C_{0}\left\vert 0\right\rangle _{d}\pm iC_{1}\left\vert
1\right\rangle _{d}),
\end{equation}%
and a phase shift of $\pi /2$ ($3\pi /2$) cancels out the phase $i$ ($-i$),
corresponding to the state $|\Psi ^{(+)}\rangle _{a,b}$ ($|\Psi
^{(-)}\rangle _{a,b}$). This step completes our protocol for splitting the
quantum information via W states. Since we cannot distinguish between $|\Phi
^{(+)}\rangle _{a,b}$ and $|\Phi ^{(-)}\rangle _{a,b}$ the success
probability for our protocol is $50\%$. In what follows we study how the
fidelity of the process described by our protocol is influenced by
nonidealities of beam splitters and photodetectors.

\emph{Losses in beam splitters and in photodetectors} - For an ideal and
symmetric beam splitter, the relationship between the input and the output
operators can be written as%
\begin{eqnarray}
a_{in}^{\dag } &\rightarrow &Ta_{out}^{\dag }+iRb_{out}^{\dag },  \label{ain}
\\
b_{in}^{\dag } &\rightarrow &Tb_{out}^{\dag }+iRa_{out}^{\dag },  \label{bin}
\end{eqnarray}%
where $a^{\dag }$ and $b^{\dag }$ are the creation operators on modes $a$
and $b$, respectively, and the coefficients $T$ and $R$ satisfy the
condition $T^{2}+R^{2}=1$. For the nonideal case, a phenomenological
operator can be added in Eqs.(\ref{ain} and \ref{bin}) in a way that the
relationship between the input and the output operators is given by \cite%
{BarnettPRA98}%
\begin{eqnarray}
a_{in}^{\dag } &\rightarrow &ta_{out}^{\dag }+irb_{out}^{\dag }+\mathfrak{L}%
_{a}^{\dag }, \\
b_{in}^{\dag } &\rightarrow &tb_{out}^{\dag }+ira_{out}^{\dag }+\mathfrak{L}%
_{b}^{\dag },
\end{eqnarray}%
where $t=\sqrt{\kappa }T$ and $r=\sqrt{\kappa }R$, with $t^{2}+r^{2}=\kappa $%
; $\mathfrak{L}_{i}$,\ $i=a,b$, are the Langevin operators accounting for
the errors introduced by the fluctuating currents within the medium
composing the $BS.$ The bosonic commutation relations for the output mode
operators lead to the requirements for the Langevin operators: $[\mathfrak{L}%
_{a},\mathfrak{L}_{a}^{\dag }]=[\mathfrak{L}_{b},\mathfrak{L}_{b}^{\dag
}]=\Gamma $ and $[\mathfrak{L}_{a},\mathfrak{L}_{b}^{\dag }]=[\mathfrak{L}%
_{b},\mathfrak{L}_{a}^{\dag }]=-\Delta $, where $\Gamma =1-\kappa $ is the
damping constant and $\Delta =0$ for symmetric $BS$s. The ground-state
expectation values for the products of pairs of Langevin operators are, for
symmetric beam splitters, $\langle \mathfrak{L}_{a}\mathfrak{L}_{a}^{\dag
}\rangle =\langle \mathfrak{L}_{b}\mathfrak{L}_{b}^{\dag }\rangle =\Gamma $, 
$\langle \mathfrak{L}_{a}\mathfrak{L}_{b}^{\dag }\rangle =\langle \mathfrak{L%
}_{b}\mathfrak{L}_{a}^{\dag }\rangle =0$, and $\langle \mathfrak{L}%
_{a}\rangle =\langle \mathfrak{L}_{a}^{\dag }\rangle =\langle \mathfrak{L}%
_{b}\rangle =\langle \mathfrak{L}_{b}^{\dag }\rangle =0$. The inefficiency
of the photodetectors can be treated in a similar manner by relating the
output operators to the input ones by \cite{SerraJOB02}%
\begin{equation}
a_{in}^{\dag }\rightarrow \sqrt{\epsilon }a_{out}^{\dag }+\mathrm{L}%
_{a}^{\dag },  \label{det}
\end{equation}%
where $\epsilon $ stands for the detector efficiency. Note that, differently
from the $BS$s, the detectors do not couple different modes \cite{SerraJOB02}%
. Besides satisfying all properties introduced above, the Langevin operators 
$\mathrm{L}_{a}^{\dag }$ also obey the commutation relations: $[\mathrm{L}%
_{a},\mathrm{L}_{a}^{\dag }]=1-\epsilon $ and $[\mathrm{L}_{a},\mathrm{L}%
_{b}^{\dag }]=0$; so the ground-state expectation values of their pair
products are: $\langle \mathrm{L}_{a}\mathrm{L}_{a}^{\dag }\rangle
=1-\epsilon $ and $\langle \mathrm{L}_{a}\mathrm{L}_{b}^{\dag }\rangle =0$.

Next, we turn to the procedure for splitting the quantum information%
, now including the loss effects. Let us begin
by considering the input state 
\begin{equation}
|\psi _{1}\rangle =(C_{0}\left\vert 0\right\rangle _{a}+C_{1}\left\vert
1\right\rangle _{a})\left\vert 1,0,0\right\rangle _{bcd}\left\vert \mathbf{0}%
\right\rangle _{R},  \label{in}
\end{equation}%
as shown in Fig. 1. Here $\left\vert \mathbf{0}\right\rangle _{R}\equiv
\prod\limits_{k}\left\vert 0\right\rangle _{k}$ stands for the state of the
environment composed of a huge number of vacuum-field states $\left\vert
0\right\rangle _{k}$. In the following, this state impinges on the four beam
splitters of the apparatus. In the first beam splitter ($BS_{1}$) the modes $%
a$ and $b$ become entangled, the whole system being described by%
\begin{align}
|\psi _{2}\rangle & =(C_{0}\left\vert 0\right\rangle _{a}+C_{1}\left\vert
1\right\rangle _{a})\left( t\left\vert 1,0,0\right\rangle _{bcd}\right.  
\notag \\
& +\left. ir\left\vert 0,1,0\right\rangle _{bcd}+\mathfrak{L}_{1}^{\dag
}\left\vert 0,0,0\right\rangle _{bcd}\right) \left\vert \mathbf{0}%
\right\rangle _{R}.
\end{align}%
After $BS_{2}$, the state $|\psi _{2}\rangle $ evolves to%
\begin{align}
|\psi _{3}\rangle & =(C_{0}\left\vert 0\right\rangle _{a}+C_{1}\left\vert
1\right\rangle _{a})\left( t\left\vert 1,0,0\right\rangle _{bcd}\right.  
\notag \\
& +\left. irt\left\vert 0,1,0\right\rangle _{bcd}-r^{2}\left\vert
0,0,1\right\rangle _{bcd}\right.   \notag \\
& +\left. ir\mathfrak{L}_{2}^{\dag }\left\vert 0,0,0\right\rangle _{bcd}+%
\mathfrak{L}_{1}^{\dag }\left\vert 0,0,0\right\rangle _{bcd}\right)
\left\vert \mathbf{0}\right\rangle _{R}.
\end{align}%
After the $BS_{3}$, and including the effects of losses in the
photodetectors via Eq.(\ref{det}), we obtain the density operator for the
whole system in the form%
\begin{eqnarray}
\rho  &=&\left\vert 0,1\right\rangle _{ab}\left\langle 0,1\right\vert |\chi
^{(01)}\rangle _{cdR}\langle \chi ^{(01)}|  \notag \\
&&+\left\vert 1,0\right\rangle _{ab}\left\langle 1,0\right\vert |\chi
^{(10)}\rangle _{cdR}\langle \chi ^{(10)}|+\rho _{res},
\end{eqnarray}%
where $|\chi ^{(01)}\rangle _{cdR}$ and $|\chi ^{(10)}\rangle _{cdR}$ are
the states corresponding to modes $c$, $d$, and reservoirs when the modes $a$
and $b$ are described by $\left\vert 0,1\right\rangle _{ab}$ and $\left\vert
1,0\right\rangle _{ab}$, respectively; $\rho _{res}$ is the residual density
operator, corresponding to the rejected terms in the detection by $D_{a}$
and $D_{b}$, i.e.,%
\begin{equation}
_{ab}\left\langle 0,1\right\vert \rho _{res}\left\vert 0,1\right\rangle
_{ab}=_{ab}\left\langle 1,0\right\vert \rho _{res}\left\vert
1,0\right\rangle _{ab}=0,
\end{equation}%
since the detection of $\left\vert 0,1\right\rangle _{ab}$ or $\left\vert
1,0\right\rangle _{ab}$ are the sole possibilities that allow us to continue
with the protocol, according to Eqs.(\ref{bell1},\ref{bell2}).

Following the ideal protocol explained above, we assume that Charlie and Bob
agree in collaborating for the reconstruction of the state. After they send
their particles to interact through the $BS_{4}$ we have the following
evolutions $\left\vert \chi ^{\left( 01\right) }\right\rangle
_{cdR}\rightarrow \left\vert \eta ^{\left( 01\right) }\right\rangle _{cdR}$
and $\left\vert \chi ^{\left( 10\right) }\right\rangle _{cdR}\rightarrow
\left\vert \eta ^{\left( 01\right) }\right\rangle _{cdR}$, where%
\begin{eqnarray}
|\eta ^{(01)}\rangle _{cdR} &=&\mathcal{N}_{01}\left\{ \left[ C_{0}\sqrt{%
\epsilon }t^{2}+2iC_{1}rt^{3}\sqrt{\epsilon }\mathrm{L}_{2}^{\dag }\right.
\right.  \notag \\
&&+iC_{1}rt^{2}\sqrt{\epsilon }\mathfrak{L}_{3}^{\dag }+C_{1}t^{3}\sqrt{%
\epsilon }\mathfrak{L}_{3}^{\dag }  \notag \\
&&-C_{1}r^{2}t\sqrt{\epsilon }\mathfrak{L}_{4}^{\dag }-iC_{1}r^{3}\sqrt{%
\epsilon }\mathfrak{L}_{4}^{\dag }  \notag \\
&&\left. -C_{1}r^{2}\sqrt{\epsilon }\mathfrak{L}_{2}^{\dag }+iC_{1}r\sqrt{%
\epsilon }\mathfrak{L}_{1}^{\dag }\right] \left\vert 0,0\right\rangle
_{cd}\left\vert \mathbf{0}\right\rangle _{R}  \notag \\
&&+\left[ C_{1}r^{4}\sqrt{\epsilon }-C_{1}r^{2}t^{2}\sqrt{\epsilon }\right]
\left\vert 1,0\right\rangle _{cd}\left\vert \mathbf{0}\right\rangle _{R} 
\notag \\
&&-\left. 2iC_{1}r^{3}t\sqrt{\epsilon }\left\vert 0,1\right\rangle
_{cd}\left\vert \mathbf{0}\right\rangle _{R}\right\} ,
\end{eqnarray}%
and%
\begin{eqnarray}
|\eta ^{(10)}\rangle _{cdR} &=&\mathcal{N}_{10}\left\{ \left[ iC_{0}rt\sqrt{%
\epsilon }+2iC_{1}rt^{3}\sqrt{\epsilon }\mathrm{L}_{1}^{\dag }\right. \right.
\notag \\
&&+C_{1}t^{3}\sqrt{\epsilon }\mathfrak{L}_{3}^{\dag }+iC_{1}rt^{2}\sqrt{%
\epsilon }\mathfrak{L}_{3}^{\dag }  \notag \\
&&+iC_{1}rt^{2}\sqrt{\epsilon }\mathfrak{L}_{4}^{\dag }-C_{1}r^{2}t\sqrt{%
\epsilon }\mathfrak{L}_{4}^{\dag }  \notag \\
&&+\left. iC_{1}rt\sqrt{\epsilon }\mathfrak{L}_{2}^{\dag }+C_{1}t\sqrt{%
\epsilon }\mathfrak{L}_{1}^{\dag }\right] \left\vert 0,0\right\rangle
_{cd}\left\vert \mathbf{0}\right\rangle _{R}  \notag \\
&&+\left[ iC_{1}rt^{3}\sqrt{\epsilon }-iC_{1}r^{3}t\sqrt{\epsilon }\right]
\left\vert 1,0\right\rangle _{cd}\left\vert \mathbf{0}\right\rangle _{R} 
\notag \\
&&-\left. 2C_{1}r^{2}t^{2}\sqrt{\epsilon }\left\vert 0,1\right\rangle
_{cd}\left\vert \mathbf{0}\right\rangle _{R}\right\} ,
\end{eqnarray}%
$\mathcal{N}_{01}$ and $\mathcal{N}_{10}$ standing for normalization. Then,
as done in the ideal protocol, a phase shift is applied on mode $d$
to change the phase according to the detected state, in a way that $%
\left\vert 0,1\right\rangle _{ab}$ ($\left\vert 1,0\right\rangle _{ab}$)
needs a phase shift of $3\pi /2$ ($\pi /2$).

Let us consider the case when Alice detects the state $\left\vert
0,1\right\rangle _{ab}$. Then the density operator is reduced to the
subsystems $c-d-R$ in the form%
\begin{equation}
\rho _{cdR}^{(01)}=|\eta ^{(01)}\rangle _{cdR}\langle \eta ^{(01)}|.
\label{r01}
\end{equation}%
Next, we trace out the reservoir in the operator above to get the fidelity $%
F_{01}=_{cd}\left\langle \Phi \right\vert \rho _{cd}^{(01)}\left\vert \Phi
\right\rangle _{cd}$, with $\rho _{cd}^{(01)}=Tr_{R}(\rho _{cdR}^{(01)})$
and $\left\vert \Phi \right\rangle _{cd}=\left\vert 0\right\rangle
_{c}(C_{0}\left\vert 0\right\rangle _{d}+C_{1}\left\vert 1\right\rangle
_{d}) $:%
\begin{eqnarray}
F_{01} &=&\mathcal{N}_{01}^{2}\left[ C_{0}^{4}\epsilon
t^{4}+2C_{0}^{2}C_{1}^{2}\epsilon
r^{3}t^{3}+4C_{0}^{2}C_{1}^{2}r^{2}t^{6}\epsilon (1-\epsilon )\right.  \notag
\\
&&+C_{0}^{2}C_{1}^{2}r^{2}t^{4}\epsilon (1-\kappa
)+C_{0}^{2}C_{1}^{2}t^{6}\epsilon (1-\kappa )  \notag \\
&&+2C_{0}^{2}C_{1}^{2}r^{3}t^{3}\epsilon +4C_{1}^{4}r^{6}t^{2}\epsilon
+C_{0}^{2}C_{1}^{2}r^{4}t^{2}\epsilon (1-\kappa )  \notag \\
&&+C_{0}^{2}C_{1}^{2}r^{6}\epsilon (1-\kappa
)+C_{0}^{2}C_{1}^{2}r^{4}\epsilon (1-\kappa )  \notag \\
&&+\left. C_{0}^{2}C_{1}^{2}r^{2}\epsilon (1-\kappa )\right] \text{,}
\end{eqnarray}%
where the normalization factor is%
\begin{eqnarray}
\mathcal{N}_{01} &=&\left[ C_{0}^{2}t^{4}\epsilon
+4C_{1}^{2}r^{2}t^{6}\epsilon (1-\epsilon )+C_{1}^{2}r^{2}t^{4}\epsilon
(1-\kappa )\right.  \notag \\
&&+C_{1}^{2}t^{6}\epsilon (1-\kappa )+4C_{1}^{2}r^{6}t^{2}\epsilon
+C_{1}^{2}r^{4}t^{2}\epsilon (1-\kappa )  \notag \\
&&+C_{1}^{2}r^{6}\epsilon (1-\kappa )+C_{1}^{2}r^{4}\epsilon (1-\kappa ) 
\notag \\
&&+\left. C_{1}^{2}r^{2}\epsilon (1-\kappa )\right] ^{-1/2}\text{.}
\end{eqnarray}

\begin{figure}[t]
\centering
\includegraphics[width=4cm]{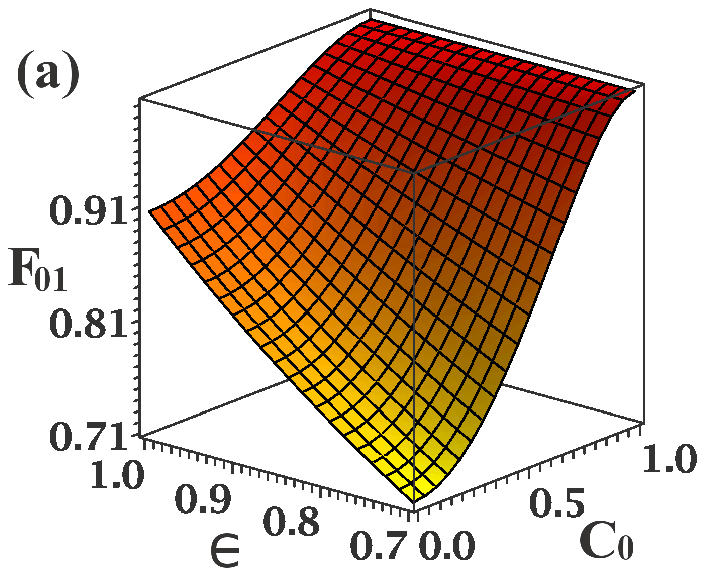} \includegraphics[width=4cm]{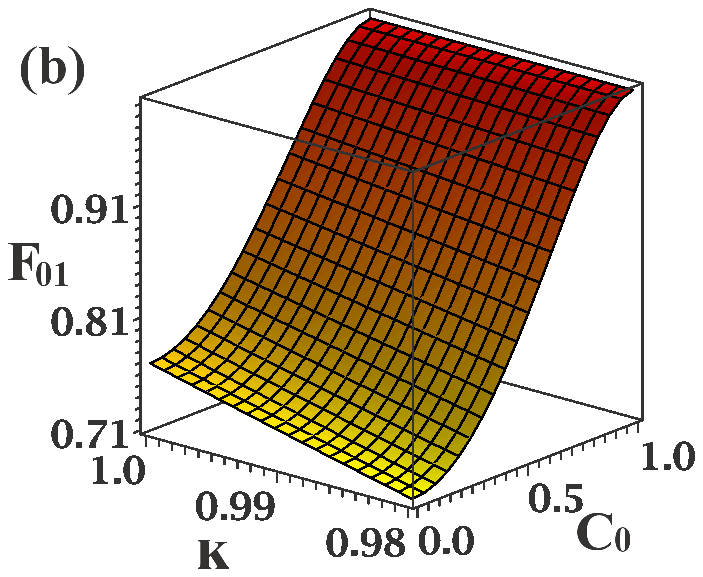}
\caption{(Color online) Fidelity of the reconstructed state considering photodetection $%
\left\vert 0,1\right\rangle _{ab}$. (a) with fixed $\protect\kappa =0.98$
(b) with fixed $\protect\epsilon =0.7$.}
\label{fig2}
\end{figure}

\begin{figure}[t]
\centering
\includegraphics[width=4cm]{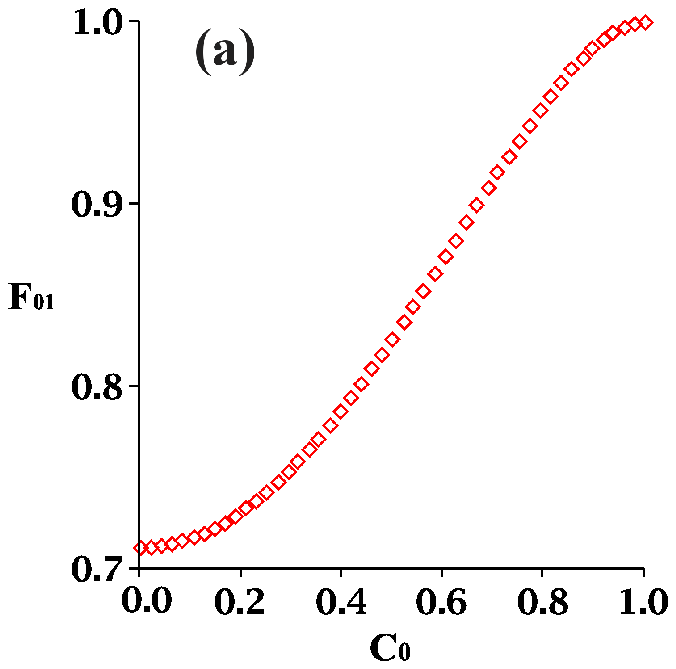} \includegraphics[width=4cm]{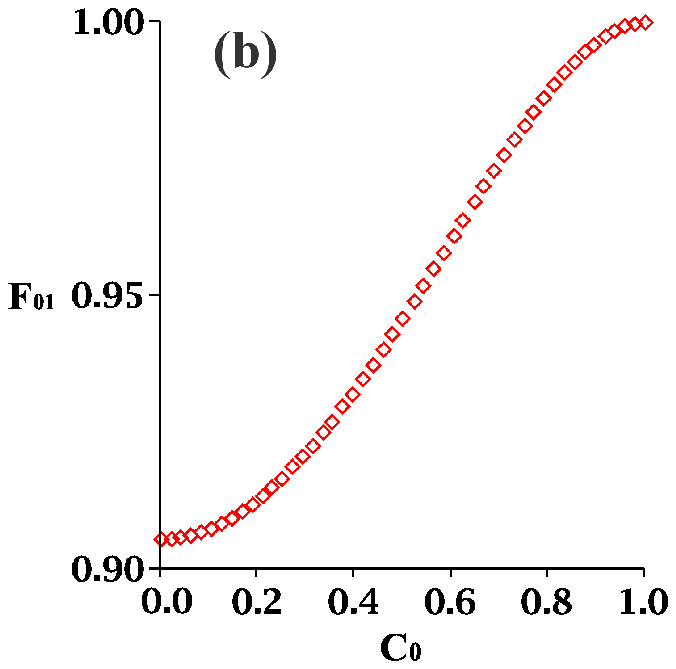}
\caption{(Color online) Fidelity of the reconstructed state considering de photodetection $%
\left\vert 0,1\right\rangle _{ab}$. In (a) we consider the worse situation
with fixed $\protect\kappa =0.98$ and $\protect\epsilon =0.7$; in (b) we
fixed $\protect\kappa =0.98$ and $\protect\epsilon =1$ (ideal detectors).}
\label{fig3}
\end{figure}

Considering symmetric $BSs$ ($50/50$), where $t_{i}=r_{i}=\sqrt{\kappa /2}$,
and current experimental parameters \cite{SerraJOB02}, e.g., $0.98\leq
\kappa \leq 1$ and $0.7\leq \epsilon \leq 1$, we find $0.72\leq F_{01}\leq 1$%
. Fig. 2a shows the fidelity $F_{01}$ for a fixed $\kappa =0.98$ and varying 
$\epsilon $ and $C_{0}$. Fig. 2b shows the same for a fixed $\epsilon =0.7$
and varying $\kappa $ and $C_{0}$ whereas Fig. 3a corresponds to the worst
case, for $\kappa =0.98$ and $\epsilon =0.7$, and varying $C_{0}$. To
clarify the relevance of the photodetectors efficiency, Fig. 3b shows the
fidelity for $\kappa =0.98$ and $\epsilon =1$ (ideal detectors).

In conclusion, we have proposed a simple and feasible scheme to
split the quantum information encoded in a Fock state superposition in
traveling waves. In addition, we show how to engineer a class of W states
suitable for perfect quantum teleportation and superdense coding, with $%
100\%$ success probability, making use of an oversimplified scheme.
Our whole scheme to split quantum information makes use of only linear
optical elements and can be accomplished through a couple of detectors, four
beam splitters, one phase shifter, and two single-photon sources. The errors
introduced by nonidealities of the beam splitters and detectors were
studied; in this case the fidelity of the state results better than that in
the classical limit, even for the worst choice of experimental parameters
concerned with the efficiency of detectors and losses in the beam splitters.

We thank the CAPES, CNPq, and FUNAPE/GO, Brazilian agencies, for the partial
supports.

\end{document}